\documentclass[prl,twocolumn,amsmath,floats,showpacs,superscriptaddress]{revtex4}

\usepackage{graphicx}
\usepackage{epsfig}
\usepackage{dcolumn}
\input{epsf} 
\begin{document}
\title{Electron-vibration interaction in single-molecule junctions: from
contact to tunneling regime}

\author{O. Tal}
\affiliation{Kamerlingh Onnes Laboratory, Leiden University, P.O.
Box 9504, 2300 RA Leiden, The Netherlands}

\author{M. Krieger}
\affiliation{Kamerlingh Onnes Laboratory, Leiden University, P.O.
Box 9504, 2300 RA Leiden, The Netherlands} \affiliation{Institute
of Applied Physics, University of Erlangen-N\"{u}rnberg,
Staudtstrasse 7/A3, D-91058 Erlangen, Germany}

\author{B. Leerink}
\affiliation{Kamerlingh Onnes Laboratory, Leiden University, P.O.
Box 9504, 2300 RA Leiden, The Netherlands}

\author{J.M. van Ruitenbeek}
\affiliation{Kamerlingh Onnes Laboratory, Leiden University, P.O.
Box 9504, 2300 RA Leiden, The Netherlands}

\begin{abstract}
Point contact spectroscopy on a H$_{2}$O molecule bridging Pt
electrodes reveals a clear crossover between enhancement and
reduction of the conductance due to electron-vibration
interaction. As single channel models predict such a crossover at
transmission probability of $\tau$$=$0.5, we used shot noise
measurements to analyze the transmission and observed at least two
channels across the junction where the dominant channel has
$\tau$$=$0.51$\pm$0.01 transmission probability at the crossover
conductance, which is consistent with the predictions for
single-channel models.
\end{abstract}

\date{\today}
 \maketitle


    A molecule bridging between two metallic electrodes
provides the opportunity to explore  the interactions between
mechanical motion (molecular vibrations) and electron transport at
the atomic-scale. The influence of a vibration mode on the
conductance of such junctions is measured by inelastic electron
tunnelling spectroscopy (IETS)~\cite{Stipe1998,Park2000} or by
point contact spectroscopy (PCS)~\cite{Agrait2002a,Smit2002}. Both
spectroscopies were originally developed for macroscopic
junctions. IETS was first investigated for molecules buried inside
a metal-oxide-metal tunneling junction~\cite{Jaklevic1966} and was
later applied to single-molecule junctions using scanning
tunnelling microscopy (STM)~\cite{Stipe1998}. PCS was first
investigated for the study of electron-phonon interactions in
metal wires with a sub-micron size constriction~\cite{Yanson1974},
and it was latter applied to single atom~\cite{Agrait2002a} and
molecule junctions~\cite{Smit2002} formed by mechanically
controlled break junctions (MCBJ). These techniques provide
information on the presence of the molecule~\cite{Qiu2004,Yu2004},
its structure~\cite{Smit2002}, the molecule
orientation~\cite{Djukic2005}, and the molecule-leads
coupling~\cite{Osorio2007}. Essentially IETS and PCS are
associated with a similar measurement of current (or its first and
second derivatives) as a function of voltage between the two leads
but operate in the opposite limits of low conductance
($G\ll1$G$_0$ where G$_0=2e^{2}/h$ is the conductance quantum) for
IETS~\cite{Stipe1998}, and conductance close to 1G$_0$ for
PCS~\cite{Smit2002}, respectively.


    In off-resonance~\cite{Troisi2006} IETS and PCS measurements,
above a certain voltage bias the incoming electrons have
sufficient energy to scatter inelastically by exciting a vibration
mode at the junction. Interestingly, electron-vibration
interaction leads to an increase in the junction conductance for
junctions in the tunnelling regime (e.g., IETS done by
STM~\cite{Stipe1998}), however it decreases the conductance for
junctions in the contact regime (e.g., PCS across a Pt/H$_2$
junction~\cite{Smit2002}). The conductance enhancement in the
first case is commonly explained by the opening of an additional
tunnelling channel for electrons that lost energy to a vibration
mode~\cite{Jaklevic1966}. The conductance suppression in the
second case has been explained in the limit of perfect electron
transmission probability ($\tau$$=$1) by backscattering of
electrons that lose energy to a vibration mode and are then
restricted by Fermi statistics to taking on the opposite momentum
since at $\tau$$=$1 the forward momentum states are fully occupied
at the reduced energy~\cite{Agrait2002b}.

    For weak electron-vibration interaction, the effect of vibration
excitation on the conductance is determined merely by the
transmission probability across the junction, when using models
based on the lowest order expansion~\cite{Galperin2004} of the
electron-vibration
coupling~\cite{Paulsson2005,Viljas2005,delaVega2006} and for
symmetric coupling of the molecule to both
leads~\cite{Paulsson2007Cond,Egger2007Cond}. In this framework,
which is different from the simplified view presented above, a
combined picture for the two limits (tunnelling and contact) was
suggested by several single-level
models~\cite{Paulsson2005,delaVega2006,Paulsson2007Cond,Egger2007Cond}.
The models predict conductance enhancement below transmission
probability of $\tau$$=$0.5 and suppression of conductance above
this, due to two opposite contributions to the conductance by the
electron-vibration interaction: an inelastic scattering process
that increases the conductance and an elastic process, where a
virtual phonon is emitted and reabsorbed by the electron. The
latter effect reduces the conductance~\cite{Viljas2005}. In spite
of the theoretical efforts invested in exploring the different
regimes of the electron-vibration interactions in atomic and
molecular junctions, this issue has not been addressed
experimentally.

\begin{figure}[t!]
\begin{center}
\includegraphics[width=8.0cm]{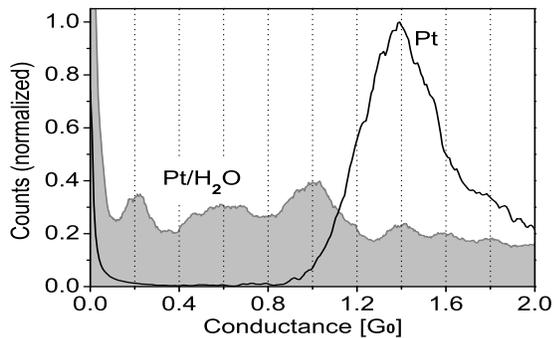}
\end{center}
\caption{Conductance histograms (normalized to the area under the
curves and set to 1 at the Pt peak) for a Pt junction (black
curve), and for Pt after introducing H$_{2}$O (filled curve). Each
conductance histogram is constructed from 1000 conductance traces
recorded with a bias of 0.2V during repeated breaking of the
contact.} \label{fig1}
\end{figure}

    In this letter we present PCS and shot noise measurements across a
single-molecule break junction formed by Pt electrodes and
H$_{2}$O molecules. By altering the electrode distance, we have
measured the effect of the electron-vibration interaction on the
differential conductance ($dI/dV$) in the transition between
tunneling and contact regimes~\cite{Frederiksen2007}. The main
transmission channels across the junctions and their probabilities
where determined allowing comparison with single-channel models
that ascribe changes in the electron-vibration interaction to the
value of the transmission probability. Our findings provide
experimental support for these models and expand their
implications to junctions involving multiple channels.

   The Pt/H$_2$O molecular junctions were formed using an MCBJ
setup~\cite{Agrait2002a} at about 5K. Clean Pt electrode apexes
are formed under cryogenic vacuum conditions by breaking a notched
Pt wire (poly-crystalline, 0.1mm diameter, 99.99\% purity). The
wire was broken by mechanical bending of a flexible substrate to
which the wire was attached. The inter-electrode distance can be
accurately adjusted (with sub-atomic precision) by fine bending of
the substrate using a piezo-element. The formation of a clean Pt
contact is verified by conductance histograms made from 1000
conductance traces taken during repeated contact stretching as
presented in Fig. 1 (black curve). The single peak around 1.4G$_0$
provides a fingerprint of a clean Pt contact~\cite{Smit2002}.

    Deionized H$_2$O~\cite{milliq} was placed in a quartz tube and
was degassed by four cycles of freezing, pumping and thawing.
While the Pt junction was broken and formed repeatedly, H$_{2}$O
molecules were introduced to the junction through a heated
capillary (baked-out prior to cooling). The junction exposure to
H$_{2}$O is controlled by a leak-valve at the top of the capillary
and by the capillary temperature. Following water introduction the
typical Pt peak in the conductance histogram is suppressed and
contributions from a wide conductance range are detected (see Fig.
1 filled curve) with minor peaks around 0.2, 0.6, and 1.0 G$_0$
(peaks around 0.95, and 1.10 G$_0$ are sometimes observed as
well). The continuum in the conductance counts implies a variety
of stable junction configurations that we exploit for spectroscopy
measurement on junctions with different conductance as discussed
next.

\begin{figure}[t!]
\begin{center}
\includegraphics[width=8.0cm]{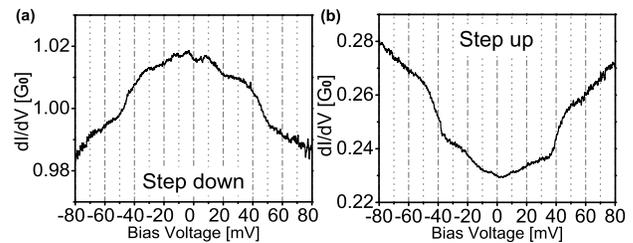}
\end{center}
\caption{Differential conductance ($dI/dV$) as a function of the
bias voltage for two different Pt-H$_2$O-Pt junctions with
zero-bias conductance of 1.02$\pm$0.01G$_0$ (a) and
0.23$\pm$0.01G$_0$ (b). Above a certain bias voltage the energy of
the incoming electrons exceeds the energy of a molecular vibration
mode and some of the electrons (a few percents) lose energy by
exciting the vibration mode. Consequently the conductance drops
("step down", a) or is enhanced ("step up", b) by the
electron-vibration scattering.} \label{fig2}
\end{figure}

    Figure 2 presents differential conductance measurements as a
function of voltage across the Pt/H$_2$O junction at two different
zero-voltage conductance values: 1.02$\pm$0.01G$_0$ (a) and
0.23$\pm$0.01G$_0$ (b). Junctions with different zero-bias
conductance are formed by altering the distance between the Pt
contacts or by re-adjusting a new contact. The steps in the
conductance that appear at 46mV in Fig. 2 (a), and 42mV in Fig. 2
(b) indicate the onset of a vibration excitation at these voltages
(the origin of the steps as due to the electron-vibration
interaction was verified by isotope substitution, e.g.
~\cite{Stipe1998,Djukic2005}). Vibration modes around 42meV are
typical for Pt/H$_2$O junctions and may be associated with a
rotation mode~\cite{note42meV}. While in (a) the differential
conductance is decreased ("step-down"), the curve (b) taken at
lower zero-voltage conductance shows an increase in the
differential conductance ("step-up"). These two examples
demonstrate that both conductance suppression and enhancement can
be observed at a relatively high conductance (much higher then the
typical tunneling conductance).

    Collecting many $dI/dV$ spectra at different zero-voltage
conductance values allow us to focus on the transition between the
two cases. Figure 3 presents the distribution of differential
conductance curves with step-up (grey) and step-down (dark)
according to their zero-voltage conductance. Curves with step-up
appear below 0.57$\pm$0.03G$_0$ and curves with step-down where
detected only above 0.72$\pm$0.03G$_0$. Thus the crossover between
conductance enhancement to conductance reduction by the
electron-vibration interaction occurs between these two values.

    According to the single-channel models the crossover is
expected at a transmission probability of $\tau$$=$0.5
($\tau$$<$0.5) for junctions with similar (different) coupling to
the electrodes and in any case not higher then
$\tau$=0.5~\cite{Paulsson2007Cond,Egger2007Cond}. The measured
conductance at the crossover is above 0.5G$_0$. However, more then
one channel can contribute to the measured conductance as
demonstrated by ${\rm Landauer's}$
formula~\cite{Landauer1957,Landauer1970}: $G=G_0\sum_{i}\tau_{i}$,
where $\tau_{i}$ is the transmission probability of the $i-th$
channel across the junction. In order to examine our findings in
view of the theoretical predictions we determined the number of
transmission channels and their probability using shot noise
measurements.

\begin{figure}[t!]
\begin{center}
\includegraphics[width=8.0cm]{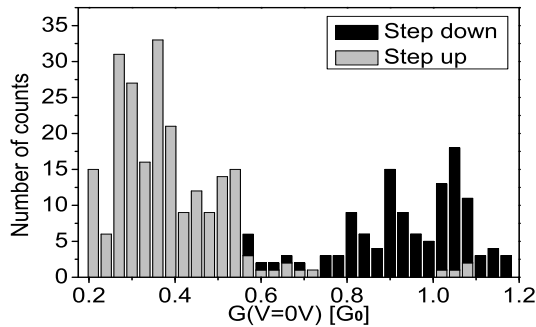}
\end{center}
\caption{Histogram of step-up (grey) and step-down (dark) features
in dI/dV spectra for Pt/H$_2$O junction as a function of zero-bias
conductance. A crossover is observed between 0.57 and
0.72$\pm$0.03G$_0$.} \label{fig3}
\end{figure}

    Shot noise results from time-dependent fluctuations in
the electrical current caused by the discreteness of the electron
charge. When electrons flow across a point contact (e.g. a single
atom or molecule junction), the noise level is determined by the
number of available transmission channels across the junction and
their transmission probabilities, ${\tau_{i}}$. The total noise
level of a quantum point contact for temperature T and applied
bias voltage V is given by~\cite{Blanter2000}:
\begin{equation}S_I=2eV{\rm coth}\left(\frac{eV}{2kT}\right)\frac{2e^{2}}{h}
\sum_{i}\tau_{i}(1-\tau_{i})+4kT\frac{2e^{2}}{h}\sum_{i}\tau_{i}^{2}\end{equation}
where k is Boltzmann's constant. Thus in combination with
Landauer's equation the main transmission probabilities can be
resolved from noise and conductance measurements.

    We have measured noise using the method described in Ref.~\cite{Djukic2006}.
Once a stable contact was established at a certain conductance,
the noise power was measured as a function of frequency at
different current bias values, where at each bias 10,000 noise
spectra where averaged. $dI/dV$ spectra were measured before and
after every set of noise measurements to verify that the same
contact was maintained during the measurements, and to avoid
junctions with considerable $dI/dV$ fluctuation within the
measurement bias range (conductance fluctuations are ascribed to
electron interference due to scattering centers in the vicinity of
the junction~\cite{Ludoph2000}). The noise at non-zero bias is
composed from thermal and shot noise (see Eq.(1); both are white
noise in the measured frequency range of 0-100KHz) and
1/f-noise~\cite{Dutta1981}. Since the noise signal is suppressed
at high frequencies due to the low-pass characteristic of the
measurement setup, the data was corrected for the transfer
characteristics obtained from the thermal noise which was measured
at zero bias~\cite{Djukic2006}. The 1/f noise contribution was
identified by its dependence on $V^{2}$ (unlike the shot noise
dependence on $V$) and was removed from the curves taken at
non-zero bias. Finally, the thermal noise is removed by
subtraction of the curve taken at zero bias from the rest of the
curves taken at different finite biases.

\begin{figure}[t!]
\begin{center}
\includegraphics[width=8.0cm]{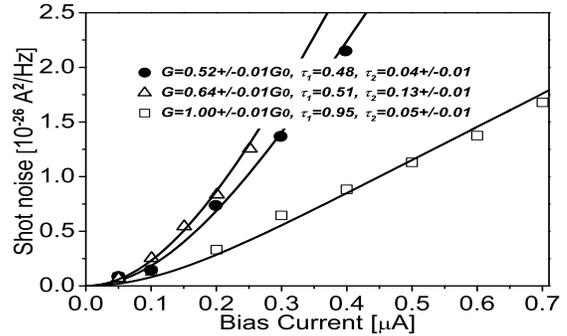}
\end{center}
\caption{Shot noise as a function of the bias current. The symbols
present shot noise measured on contacts with G=0.52$\pm$0.01G$_0$
(bullets), G=0.64$\pm$0.01G$_0$ (open triangles), and
G=1.00$\pm$0.01G$_0$ (open squares). Fitting the data with Eq.(1)
(solid curves) gives the decomposition of the total conductance in
terms of the conduction channels ($\tau_{1}$ and $\tau_{2}$).}
\label{fig4}
\end{figure}

\begin{table}[b!]
\begin{center}
\begin{tabular}{|p{1.2cm}|p{1.0cm}|p{1.0cm}|p{1.0cm}|p{1.0cm}|p{1.0cm}|p{0.8cm}|}
 \hline
 $G [G_o],\tau$ & 0.52 & 0.62 & 0.64 & 0.96 & 1.00 & $\pm0.01$ \\ \hline
 $\tau_{1}$ & $0.48$ & $0.51$ & $0.51$ & $0.93$ & $0.95$ & $\pm0.01$ \\ \hline
 $\tau_{2}$ & $0.04$ & $0.11$ & $0.13$ & $0.03$ & $0.05$ & $\pm0.01$ \\ \hline
 $\tau_{1}/\tau\ [\%]$ & $92\pm2$ & $82\pm2$ & $80\pm2$ & $97\pm1$ & $95\pm1$ & $ $ \\
 \hline
\end{tabular}
\end{center}
\caption{\small{Zero bias conductance, total transmission
probability ($\tau$), main transmission probabilities ($\tau_{1} $
and $\tau_{2}$), and the ratio between the main transmission
probability and the total transmission for different Pt/H$_2$O
junctions.}} \label{table}
\end{table}

   Following this analysis several sets of shot noise as a function of current
bias were obtained for junctions with different zero bias
conductance. Figure 4 presents three examples for such data taken
on junctions with zero bias conductance of 0.52 (bullets), 0.64
(open triangles), and 1.00$\pm$0.01G$_0$ (open squares) zero-bias
conductance. The transmission probability of the main channels can
be determined by fitting Eq. (1) to the measured noise and using
Landauer's equation to obtain the total transmission probabilities
from the measured conductance. Since the fitting is extremely
sensitive to the number of channels and their
probabilities~\cite{vandenBrom1999,Djukic2006}, the freedom in
choosing the main transmission probability is limited, in this
case to $\pm$0.01, while choosing more then two channels is
restricted to small additional channels that do not affect the
main probability (in the range of $\pm$0.01).

    The main transmission probabilities obtained for junctions with
different conductance are presented in Table~ I. The reliability
of the noise measurements is demonstrated by the consistency of
the transmission probabilities between independent measurements
done on different junctions with relatively close conductance
(e.g. 0.62 and 0.64$\pm$0.01G$_0$ or 0.96 and 1.00$\pm$0.01G$_0$).
Keeping in mind the prediction for inversion in the
electron-vibration effect at a single transmission probability of
0.5, it is interesting to examine the main transmission
probability at different conductance values. At 0.52G$_0$ the main
transmission probability is lower then 0.5 whereas at 0.96G$_0$ it
is well above 0.5. The main transmission probability crosses
0.51$\pm$0.01 at conductance of 0.62-0.64G$_0$.

    Considering both PCS and shot noise measurements we observe
a clear crossover between enhancement of the differential
conductance to suppression at 0.57-0.72G$_0$. It is found that all
the examined junctions have one transmission channel which is
dominant over the other channel(s) (forth row of Table~I).
Finally, the transmission probability of the dominant channel
crosses $\tau$$=$0.51$\pm$0.01 at a zero-bias conductance of
0.62-0.64G$_0$, right at the center of the conductance range where
the crossover between differential conductance enhancement to
suppression takes place. The agreement of these findings with the
single channel models that predict a transition at $\tau$$=$0.5
provided that the molecule coupling to both electrodes is similar
suggests that the latter condition is fulfilled, and that the
conductance suppression or enhancement by the electron-vibration
interaction is determined by the value of the \textit{dominant}
transmission probability. In more general perspective the lowest
order expansion for the electron-vibration interaction correctly
predicts a crossover in sign of the step in differential
conductance at transmission of $\tau$$=$0.5. Even in the presence
of additional conductance channels this effect can be observed for
the dominant channel, as in the case for our Pt/H$_2$O system.

    When there is not a single dominant channel (as was observed
for Pt/C$_6$H$_6$(benzene) junctions~\cite{unpablished}), no clear
transition between conductance enhancement to suppression is
observed. However many other junctions follow the general behavior
of step-down near 1G$_0$ (e.g, Au atomic wires~\cite{Agrait2002a},
Pt/H$_2$~\cite{Smit2002}), and step-up below~0.3G$_0$ (e.g, Ag
atomic-wires decorated with oxygen~\cite{ThijssenCondMat}).

    This work is part of the research program of the
``Stichting FOM'', which is financially supported by NWO. OT is
also grateful for the AVS-Welch scholarship. MK greatly
acknowlegdes the support by the European Commission (RTN DIENOW).
We thank A. Nitzan, J. C. Cuevas and A. Levy Yeyati for fruitful
discussions.


\begin{thebibliography}{29}
\bibitem[{not()}]{note42meV}
\bibinfo{note}{This mode is insensitivite to contact stretching and to
  variations in the junction contuctance. More definit identification requires
  detailed comparison with calculations.}
\end{thebibliography}
\end{document}